\documentclass[prx,aps,twocolumn,superscriptaddress,longbibliography]{revtex4-1}

\usepackage[dvips]{graphicx}
\usepackage{amssymb,amsfonts,amsmath}
\usepackage{color}
\usepackage{ulem}
\usepackage[hidelinks]{hyperref}

\newcommand{\rv}{{\mathbf r}}
\newcommand{\ev}{{\mathbf e}}

\newcommand{\Jv}{{\bf J}}

\newcommand{\fv}{{\bf f}}

\newcommand{\vel}{{\bf v}}

\begin{document}

\title{Flow and structure in nonequilibrium Brownian many-body systems}

\author{Daniel de las Heras}
\affiliation{Theoretische Physik II, Physikalisches Institut, 
  Universit{\"a}t Bayreuth, D-95440 Bayreuth, Germany}

\author{Matthias Schmidt}
\affiliation{Theoretische Physik II, Physikalisches Institut, 
  Universit{\"a}t Bayreuth, D-95440 Bayreuth, Germany}

\date{19 December 2019, revised version: 7 May 2020., to appear in Phys.\ Rev.\ Lett.}

\begin{abstract}
We present a fundamental classification of forces relevant in
nonequilibrium structure formation under collective flow in Brownian
many-body systems.  The internal one-body force field is
systematically split into contributions relevant for the spatial
structure and for the coupled motion. We demonstrate that both
contributions can be obtained straightforwardly in computer
simulations, and present a power functional theory that describes all
types of forces quantitatively. Our conclusions and methods are
relevant for flow in inertial systems, such as molecular liquids and
granular media.
\end{abstract}

\maketitle

Nonequilibrium phenomenology in colloidal systems is both diverse and poorly understood.
Examples include lane~\cite{PhysRevE.65.021402} and band~\cite{PhysRevLett.106.228303} formation in oppositely driven
colloids,
the motility induced phase separation in active systems \cite{doi:10.1146/annurev-conmatphys-031214-014710,PhysRevLett.121.098003}, the migration
of colloidal particles induced by shear fields~\cite{lyon_leal_1998,frank_anderson_weeks_morris_2003}, and magnetically controlled dynamical self-assembly \cite{snezhko2011magnetic}.
Even the universal
processes of vitrification~\cite{golde2016correlation} and crystallization~\cite{tan2014visualizing}
are still not well understood.

Our inability to properly describe nonequilibrium phenomena
can be traced back to our lack of understanding of the internal force field, that is, the
one-body field that originates from a position- and time-resolved average of the interparticle interactions.
In equilibrium, the internal force field
depends only on the density distribution and it is well described by the widely used framework
of density functional theory (DFT)~\cite{Evans1979}.
In contrast, in non-equilibrium, the internal force field depends on both the density and the flow~\cite{PFT,custom}.
We show here that the nonequilibrium internal force field naturally splits into four fundamentally different contributions 
and we provide a method to measure each of them using computer simulations. The classification of the internal forces is based on
the direction of the force and, crucially, on whether the force acts on the particle flow or on the structure of the fluid.
We show how flow and spatial structure, which are often treated separately, naturally
interplay in generating the internal force field and, therefore, provide
a unifying framework to treat both aspects on equal footing.
Finally, we present a microscopic theory, based on the power functional~\cite{PFT}, that predicts
quantitatively all occurring types of contributions to the internal force field across a range
of fundamentally different nonequilibrium situations.

Consider a nonequilibrium overdamped Brownian system with
no hydrodynamic interactions. The time evolution 
is given by the exact one-body equation of motion
\begin{align}
  \gamma \vel(\rv,t) &= \fv_{\rm tot}(\rv,t),
  \label{EQofMotion}
\end{align}
and the continuity equation
\begin{align}
  \frac{\partial}{\partial t} \rho(\rv,t) = -\nabla\cdot\Jv(\rv,t),
  \label{EQcontinuityEquation}
\end{align}
where $\gamma$ is the single-particle friction constant against the
(implicit) solvent, $\vel(\rv,t)$ is the velocity field at position
$\rv$ and time $t$,
$\fv_{\rm tot}(\rv,t)$ is the total force field,
$\rho(\rv,t)$ is the density distribution, and $\Jv=\rho\vel$ is the
current profile. The total force comprises three contributions,
\begin{align}
  \fv_{\rm tot} &= \fv_{\rm ext} + \fv_{\rm int} + \fv_{\rm id},
  \label{EQtotalForceSplitting}
\end{align}
with the imposed external force field $\fv_{\rm ext}$, the internal force
field $\fv_{\rm int}$, and the ideal-gas diffusion $\fv_{\rm id}=-k_BT\nabla\ln\rho$;
here $k_B$ is the Boltzmann constant and $T$ is absolute temperature. 
All one-body fields above are well-defined as statistical averages of 
microscopic operators, see e.g.~\cite{custom}. The underlying many-body
dynamics are given, equivalently, by a Fokker-Planck (Smoluchowski) equation for the
probability distribution or in the Langevin picture of stochastic trajectories.

The internal force consists of adiabatic and superadiabatic
contributions $\fv_{\rm int}=\fv_{\rm ad}+\fv_{\rm sup}$~\cite{PFT,PRLandrea}.
The equilibrium-like adiabatic part, $\fv_{\rm ad}$,
is the only internal force that enters in the widespread dynamical density functional theory~\cite{Marconi1999}
and it describes the 
internal forces in a hypothetical equilibrium system (vanishing current) with the same density distribution
as the actual nonequilibrium system.
In contrast, the superadiabatic force field, $\fv_{\rm sup}$,
is of purely out-of-equilibrium origin.
Both $\fv_{\rm ad}$ and $\fv_{\rm sup}$ can be written as functional
derivatives using DFT~\cite{Evans1979} and power functional theory (PFT)~\cite{PFT}, 
respectively.

Although the continuity equation~\eqref{EQcontinuityEquation} links the flow $\vel$ and the density profile $\rho$,
there is much freedom in choosing both fields separately.
PFT assures that a mapping from $\rho$ and $\vel$ to the external force exists~\cite{PFT,custom}.
Hence $\rho$ and $\vel$ constitute genuine variables rather than being the result of a prescribed
driving mechanism.
It is possible to find e.g. two
systems that share the same density profile, but have different velocity profiles. A simple
situation consists of a family of flows differing only in magnitude of the velocity~\cite{custom}.
However, much more complex cases exist, since the continuity equation~\eqref{EQcontinuityEquation}
does not impose any restriction 
on the curl of the current. Hence the addition of any divergence--free vector field
to the current keeps the continuity equation satisfied with identical density profile.
There exist also systems that share the same flow but posses different density profiles.
It is therefore natural to split the equation of motion~\eqref{EQofMotion} into
two interrelated equations, one for the flow and one for the structure, given respectively by
\begin{align}
\gamma\vel &= \fv_{\rm flow}+\fv_{\rm ext,f},\label{eqFlow}\\
0 &=\fv_{\rm id}+\fv_{\rm ad}+\fv_{\rm str}+\fv_{\rm ext,s}.\label{eqStr}
\end{align}
The diffusive term, $\fv_{\rm id}$, and the adiabatic force, $\fv_{\rm ad}$, depend only on
the density profile,
and therefore act on the structure~\eqref{eqStr}. The external force can
act on both structure and flow and therefore it splits into a contribution that acts on the flow, 
$\fv_{\rm ext,f}$, and one that acts on the structure, $\fv_{\rm ext,s}$, such that $\fv_{\rm ext}=\fv_{\rm ext,f}+\fv_{\rm ext,s}$. The 
total superadiabatic force field, $\fv_{\rm sup}$,
also acts on both flow and structure and hence can
be split according to
\begin{equation}
\fv_{\rm sup}=\fv_{\rm flow}+\fv_{\rm str}.\label{EQsupf}
\end{equation}
Equations \eqref{eqFlow} and \eqref{eqStr} are coupled since $\fv_{\rm flow}$ and
$\fv_{\rm str}$ are functionals of both $\rho$ and $\vel$. However, both equations
describe different phenomena; the velocity field is determined by the flow, Eq.~\eqref{eqFlow}, whereas the 
density profile is given by the structural force balance \eqref{eqStr}. As we demonstrate below, both 
superadiabatic forces behave differently under motion reversal.

\begin{figure}
\centering
\includegraphics[width=1.00\columnwidth]{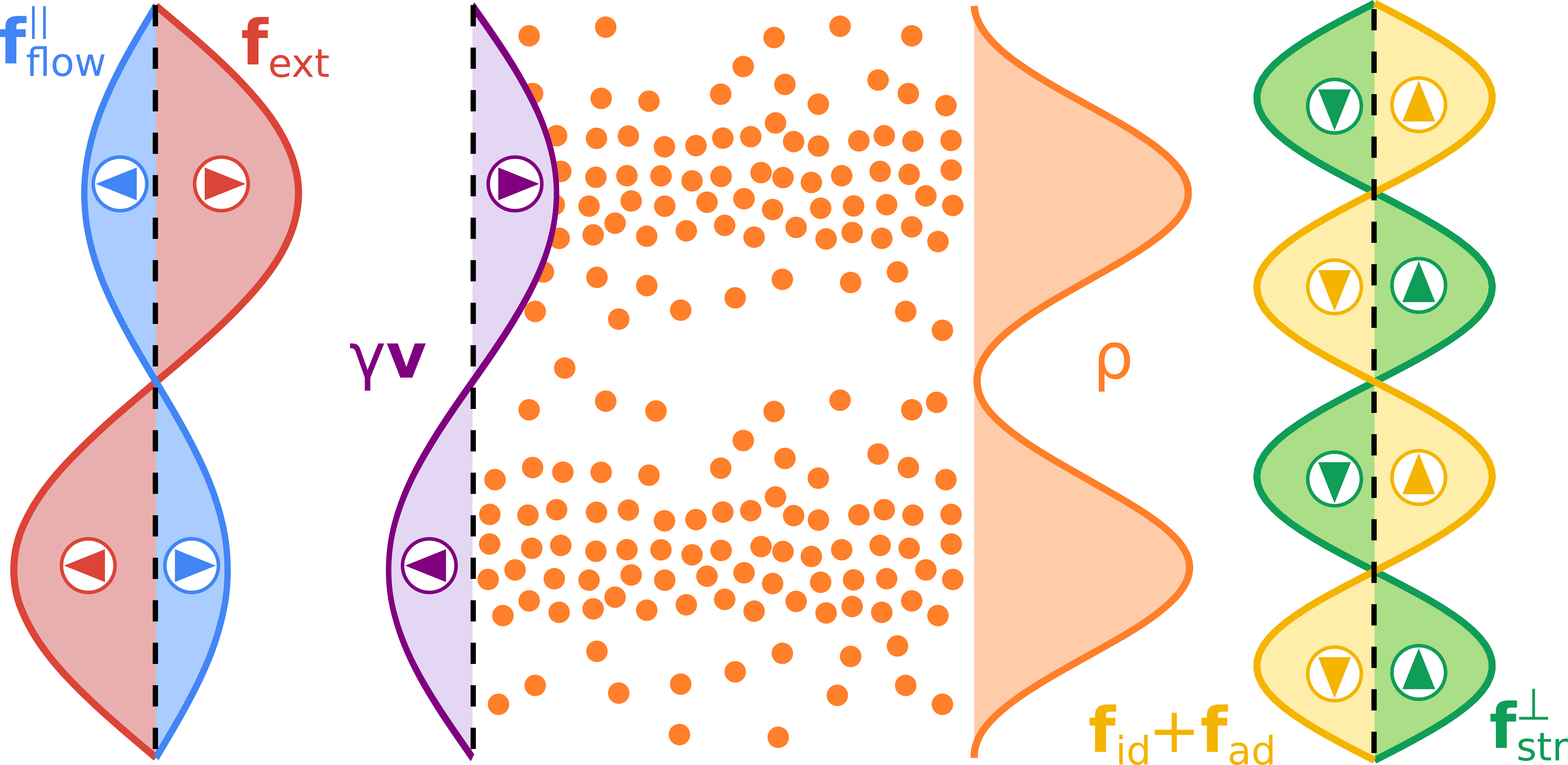}
\caption{Force balance in Kolmogorov flow. A colloidal system is driven
with a sinusoidal external force $\fv_{\rm ext}$ (red). The force creates
a flow $\gamma\vel$ (violet) that generates a superadiabatic flow force $\fv_{\rm flow}^{||}$ (blue)
of viscous nature 
and a density modulation $\rho$ (orange) which is the result of the force balance
between the superadiabatic structural, $\fv_{\rm str}^{\perp}$ (green) and the sum of adiabatic $\fv_{\rm ad}$ and
diffusive $\fv_{\rm id}$ (yellow) forces. The arrows indicate the directions of the forces.
An illustrative microstate is shown with particles depicted as orange circles.
}
\label{fig1}
\end{figure}

To gain insight into the physical properties of~\eqref{EQsupf},
we address first the force balance of the well-known phenomenon of shear migration~\cite{shearmigration,shearmigration1}.
A broad range of further types of drivings are analysed below. For simplicity we consider steady states, where all one-body quantities are
time independent and hence $\nabla\cdot\Jv=0$, cf.~\eqref{EQcontinuityEquation}. 

Consider
a colloidal system undergoing a Kolmogorov-like flow~\cite{kolmogorov}, see Fig.~\ref{fig1}. The particles are
driven by a sinusoidal external field $\fv_{\rm ext}\propto\sin(y)\hat{\mathbf x}$ (red) which
creates a steady state with a flow $\vel$ parallel to the driving (violet) in ${x}$-direction.
In the direction perpendicular
to the flow, a density modulation $\rho(\rv)=\rho(y)$ (orange) appears since
the particles migrate towards the region of low shear rate.
The superadiabatic forces here are easy to interpret~\cite{PRLstr}.
A dissipative viscous-like force
$\fv_{\rm flow}^{||}$ 
(blue) opposes the flow. The force is anti-parallel to the flow and it clearly changes
direction if the external driving is reversed.
The density modulation in the direction perpendicular
to the flow is sustained by a nondissipative structural superadiabatic force $\fv_{\rm str}^\perp$ (green) that remains
unchanged under flow reversal and that cancels both the diffusive and the adiabatic forces (yellow).
In this geometry, the flow forces are of viscous nature and parallel to the flow, whereas
the structural forces are perpendicular to it. As both superadiabatic components are orthogonal
to each other, measuring them separately
in computer simulations is simple. However, in general we expect also 
structural forces to occur parallel to the flow, $\fv_{\rm str}^{||}$, and flow
forces to occur perpendicular to the flow, $\fv_{\rm flow}^\perp$. To split the superadiabatic
forces into all these constituents, we consider what we call ``reverse'' or ''backward'' state,
which possess the same density profile as the original ``forward'' state but it has the opposite flow.
Hence, indicating quantities in the reverse state by a prime, we have
\begin{equation}
  \rho'(\rv) = \rho(\rv),\quad\vel'(\rv) = -\vel(\rv).\label{EQreverseflow}
\end{equation}
As a direct consequence of the structure of ~\eqref{eqFlow} and~\eqref{eqStr},
the flow component of the superadiabatic force reverses it sign in the reverse state ($\fv_{\rm flow}'=-\fv_{\rm flow}$), whereas 
the structural component remains unchanged ($\fv_{\rm str}'=\fv_{\rm str}$), i.e.,
\begin{align}
\fv_{\rm sup}'=\fv_{\rm flow}'+\fv_{\rm str}'=-\fv_{\rm flow}+\fv_{\rm str}.\label{EQsupvs}
\end{align}

Once flow and structural components have been identified, we project the vector fields onto the directions parallel $(||)$
and perpendicular $(\perp)$ to the velocity field. The final and complete splitting of superadiabatic forces is
\begin{align}
\fv_{\rm sup}=\fv_{\rm flow}^{||}+\fv_{\rm flow}^{\perp}+\fv_{\rm str}^{||}+\fv_{\rm str}^{\perp}.\label{EQfull}
\end{align}

The ideal diffusion and the adiabatic forces do not change in the reverse
state since they only depend on the density profile, i.e.,
$\fv_{\rm id}' =  \fv_{{\rm id}}$ and $\fv_{\rm ad}' =  \fv_{{\rm ad}}$.
Therefore, using ~\eqref{EQofMotion} and \eqref{EQsupvs}, we arrive at the following equations of motion in the forward
and in the reverse states, respectively,
\begin{align}
\gamma \vel &= \fv_{{\rm id}} + \fv_{{\rm ad}} +  \fv_{{\rm flow}} + \fv_{{\rm str}} + \fv_{{\rm ext}},\label{EQforward}\\
-\gamma \vel &= \fv_{{\rm id}} + \fv_{{\rm ad}}   -\fv_{{\rm flow}} + \fv_{{\rm str}} + \fv_{\rm ext}'.\label{EQbackward}
\end{align}
Adding and subtracting~\eqref{EQforward} and~\eqref{EQbackward} yields:
\begin{eqnarray}
  \fv_{{\rm flow}}&=&\gamma \vel-\left(\fv_{{\rm ext}}-\fv_{\rm ext}'\right)/2,\label{EQfs1}\\
  \fv_{{\rm str}}&=&-\fv_{{\rm id}}-\fv_{{\rm ad}}-\left(\fv_{{\rm ext}}+\fv_{\rm ext}'\right)/2.\label{EQfs}
\end{eqnarray}
Alternatively, from \eqref{EQsupf} and \eqref{EQsupvs} it follows that
\begin{eqnarray}
\fv_{\rm flow}&=\left(\fv_{\rm sup}-\fv_{\rm sup}'\right)/2,\label{EQsups1}\\
\fv_{\rm str}&=\left(\fv_{\rm sup}+\fv_{\rm sup}'\right)/2.\label{EQsups}
\end{eqnarray}
The final steps to apply either Eqs.~\eqref{EQfs1}-\eqref{EQfs} or~\eqref{EQsups1}-\eqref{EQsups} are (i) to find
the external force $\fv'_{\rm ext}$ that reverses the flow and (ii) to split
the internal force field, $\fv_{\rm int}$, into adiabatic and superadiabatic contributions.
In general $\fv_{\rm ext}'\neq-\fv_{\rm ext}$ due to nonvanishing structural (flow) 
forces (parallel) perpendicular to the flow, see Eqs.~\eqref{EQforward}-\eqref{EQbackward}.

The four components of the superadiabatic forces~\eqref{EQfull} are accessible in many-body BD simulations.
In Ref.~\cite{custom} we developed an iterative method to
construct the external force field that generates a given (prescribed)
time evolution of a Brownian system; a brief description is provided
in the Supplemental Material (SM)~\cite{Supp}.
Using this ``custom flow'' method it is straightforward
to calculate in BD simulations $\fv_{{\rm ext}}'$ since
$\rho$ and $\vel$ are known, see Eq.~\eqref{EQreverseflow}.
Splitting the internal force field into adiabatic
and superadiabatic contributions is also simple~\cite{PRLandrea,custom}:
we first find the adiabatic external force $\fv_{\rm ext}^{\rm ad}$,
i.e. the conservative external force field that generates the
desired density profile in equilibrium ($\vel=0$). To this end we
use the method of Ref.~\cite{custom}.
The adiabatic force $\fv_{\rm ad}$ is then the internal force in 
presence of $\fv_{\rm ext}^{\rm ad}$. The superadiabatic force is 
the difference between the total internal and the adiabatic fields, $\fv_{\rm sup}=\fv_{\rm int}-\fv_{\rm ad}$.

Knowledge of the superadiabatic forces in the forward and in the  reverse
states gives direct access to the flow and the structural components
via~\eqref{EQsups1} and \eqref{EQsups}.
Using $\fv_{\rm ext}^{\rm ad}$ the structural force~\eqref{EQfs} can be also expressed using only 
external forces $\fv_{\rm str}=\fv_{\rm ext}^{\rm ad}-(\fv_{{\rm ext}}+\fv_{\rm ext}')/2$ since in 
the adiabatic (equilibrium) system $\fv_{\rm ext}^{\rm ad}+\fv_{\rm id}+\fv_{\rm ad}=0$.
Once the flow and the structural
components are known, we project them onto the local flow direction, given by
$\hat \ev_v(\rv)=\vel(\rv)/|\vel(\rv)|$, to obtain the parallel and perpendicular components,
\begin{equation}
\fv_\alpha^{||}=(\fv_\alpha\cdot\hat \ev_v) \hat\ev_v,\quad\fv_\alpha^{\perp}=\fv_\alpha-\fv_{\alpha}^{||},\quad\alpha=\{\rm flow,\rm str\}.
\end{equation}

\begin{figure*}
\centering
\includegraphics[width=0.90\textwidth]{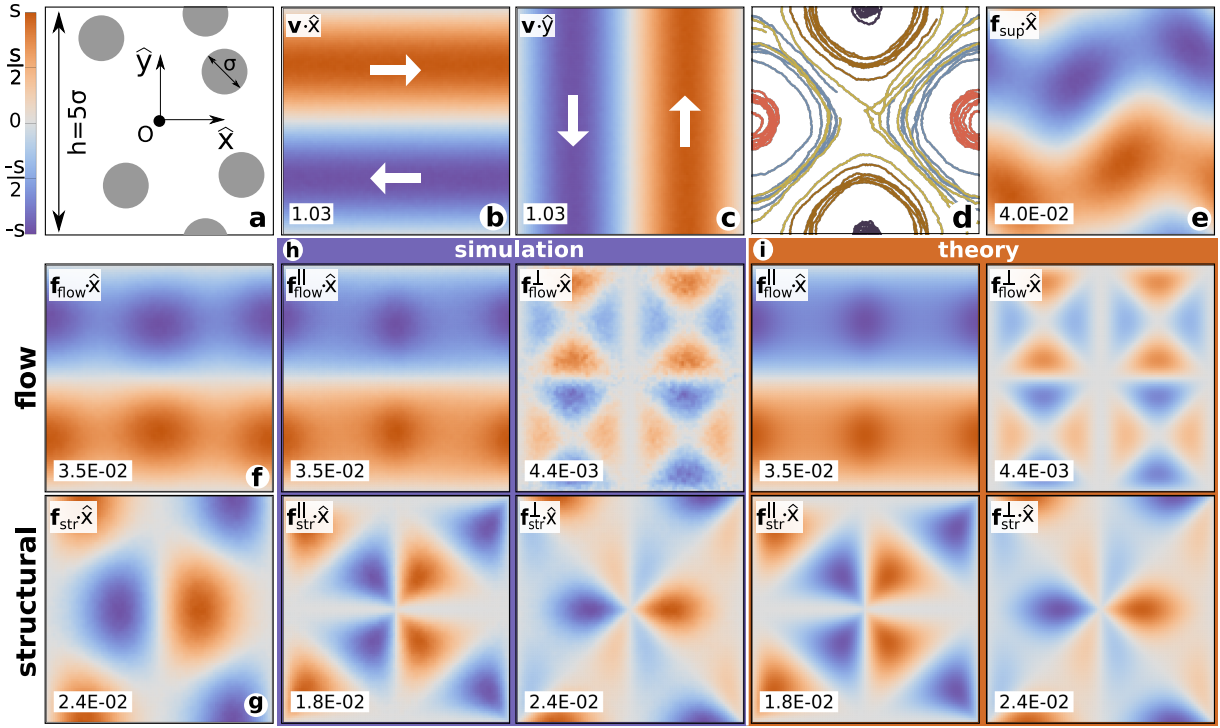}
\caption{Superadiabatic forces. (a) Schematic of the simulation box.
Sampled (b) ${x}$- and (c) ${y}$-components of the imposed steady state velocity profile.
	(d) Characteristic trajectories followed by the particles in absence of Brownian noise. 
(e) ${x}$-component of the total superadiabatic force measured in BD.
Flow (f) and structural (g) components of the superadiabatic force according to BD.
(h) Splitting of flow and structural forces into their parallel and perpendicular components, as indicated. 
Data obtained with BD simulations. (i) Same as (h) but using power functional 
theory. All forces are in units of $\epsilon/\sigma$. The flow is in units of $\tau/\sigma$. The scale factor $s$
(see color bar) is indicated in each plot. The $y$-components are depicted in the SM~\cite{Supp}.}
\label{fig2}
	\vspace{-0.3cm}
\end{figure*}

We study the fundamental splitting~\eqref{EQfull} of the superadiabatic forces in a simple two-dimensional 
system of purely repulsive particles (Weeks-Chandler-Anderson potential~\cite{doi:10.1063/1.1674820} with
$\sigma$ and $\epsilon$ as length and energy parameters, respectively). The particles are
confined in a square box of length $h$ (centered at the origin) with periodic boundary conditions, see Fig.~\ref{fig2}a.

To illustrate the classification of the superadiabatic forces, we
analyse eight different steady states. We use custom flow~\cite{custom,Supp} to
impose $\vel$ and $\rho$ and then find the corresponding external fields both
in the forward and in the reverse states. The steady states cover fundamentally different cases such as divergence-free flow,
curl-free flow, and families of flows with the same velocity profiles but different density profiles.
The description of the flows along with plots of all forces and simulation details are given in the SM~\cite{Supp}.
Here, we just illustrate the complexity and richness of the superadiabatic forces for a representative example (flow number $1$ in SM~\cite{Supp}).
The imposed steady state is a generalized Kolmogorov flow, in which each component of the velocity is a pure sinusoidal wave,
\begin{equation}
\vel(\rv) =\left( \begin{matrix}
           v_0 \sin(2\pi y /h) \\
           v_0 \sin(2\pi x /h) 
         \end{matrix}\right),\;\;\rho(\rv) =  \rho_0,\label{EQstst}
\end{equation}
with $v_0\tau/\sigma=1.0$, constant density $\rho_0=0.2\sigma^{-2}$, and $\tau=\sigma^2\gamma/\epsilon$ is the time unit.
The two Cartesian velocity components sampled in BD simulations are shown in Fig.~\ref{fig2}b and \ref{fig2}c.
Illustrative particle trajectories are shown in Fig.~\ref{fig2}d: the particles wind around specific points,
which constitute defects in the velocity field, i.e., points at which the direction of the velocity is ill-defined. To
better highlight the motion, the trajectories were calculated in absence of Brownian motion.
The imposed density profile is uniform, and therefore the ideal diffusive and the
adiabatic force fields, both gradient fields, vanish identically. The internal force field is hence purely superadiabatic. In Fig.~\ref{fig2}e we show the
${x}$-component of the superadiabatic force field. Due to the symmetry of the flow, the ${y}$-component
(shown in SM~\cite{Supp}) is simply the ${x}$-component after a ninety degrees anticlockwise rotation about the origin followed by a reflection
through the ${y}$-axis. The splitting into flow and
structural forces~\eqref{EQsupf} is shown in Figs.~\ref{fig2}f and \ref{fig2}g, respectively.
Both force fields are of the same order of magnitude but they play different roles. The flow force is mostly of viscous nature opposing the direction of motion.
The structural force is dominated by a migration-like term. It is quite complex, as it tends to
move the particles towards the defects with cyclonic vorticity (those located at the middle of the sides of the box),
away from the hyperbolic defects (center and corners of the box).
The density is, however, constant
by construction and therefore the structural 
force does not create any density modulation. Instead, the structural force is balanced by the structural component of the external field,
$\fv_{\rm ext,s}$, cf.~\eqref{eqStr}. The external force field
performs two different tasks: (i) it creates the flow and (ii) it compensates the 
structural forces that are generated as a result of the flow.

The full splitting into the four types of nonequilibrium forces~\eqref{EQfull} is shown in Fig.~\ref{fig2}h. Clearly all types of forces,
including the flow forces perpendicular to the flow and the structural forces parallel to the flow exist, and no contribution is negligible. The parallel
flow force $\fv_{\rm flow}^{||}$ can be understood as arising from viscosity. However, although the flow is made of pure sinusoidal waves, 
$\fv_{\rm flow}^{||}$ exhibits a complex spatial structure and it is not a simple sine wave opposing the flow, as one would naively
expect according to a Navier-Stokes description of the viscous force. Moreover, the perpendicular component of the flow force is highly
non-trivial and cannot be interpreted as a viscous response.

Although this example is restricted to a case of homogeneous
density, we show in the SM~\cite{Supp} a variety
of steady states with inhomogeneous density profiles. The versatility of
the custom flow method allows us to e.g.\ analyse a steady
state (flow number $2$) with the same velocity profile as in~\eqref{EQstst}
but with an inhomogeneous density profile. Comparing systems
with the same $\vel$ but different $\rho$ is useful to gain
insight into the structure of the power functional that generates the 
superadiabatic forces. 

The superadiabatic forces
are functional derivatives of the functional $P_t^{\rm exc}$ with respect to
the current~\cite{PFT,Supp}. 
The splitting into flow and structural forces
can be analyzed via a power series of the velocity
field~\cite{PRLnablaV,PRLstr}. Terms odd (even) in powers of $\vel$ lead to
structural (flow) forces that must be even (odd) in powers of $\vel$. 
Note that (i) the functional differentiation 
reduces by one the power of the velocity field, and (ii) flow (structural) forces change (do not change)
sign upon flow reversal. The simplest approximation for $P_t^{\rm exc}$ is, 
in essence, a space integral of a quadratic form in the local velocity gradient~\cite{PRLnablaV}.
The resulting superadiabatic force is a flow force (no structural terms)
that represent the viscous and shear responses in the Navier-Stokes equations. Higher order terms, 
third order in powers of the velocity field, are required to generate all the structural forces. These terms are based on spatial integrals
of rotational invariants of the type ${\sf L}_{ijkl}(\nabla_i\vel_j)(\vel_k)(\vel_l)$ and ${\sf Q}_{ijklmn}(\nabla_i\vel_j)(\nabla_k\vel_l)(\nabla_m\vel_j)$
with ${\sf L}$ and ${\sf Q}$ isotropic tensors of rank fourth and sixth, respectively.
We find it necessary to include fourth order terms to correctly describe the flow forces in some
of the steady states analysed. 
The theory reproduces from first principles the complex shape of
the superadiabatic forces computed in BD simulations (only the magnitude
of the forces needs to be adjusted)
Compare e.g. Figs. \ref{fig2}h (simulation) and \ref{fig2}i (PFT). Further comparisons
and details about $P_t^{\rm exc}$ are provided in SM \cite{Supp}. 

Although we used custom flow~\cite{custom} to prescribe the flow,
the splitting of superadiabatic forces is general and applies to the standard situation where $\fv_{\rm ext}$
is prescribed instead of the flow itself. Custom flow is then required
only to obtain the reverse state. The custom flow method also works in time-dependent nonequilibrium situations~\cite{custom}.
Hence the splitting of the superadiabatic forces can be done in any situation.
Our analysis is devoted to overdamped Brownian systems without hydrodynamic interactions.
Therefore, the superadiabatic forces are solely generated by the interparticle interactions in nonequilibrium.
Using an inert solvent has allowed us to isolate the superadiabatic effects from those due to hydrodynamics.
Inclusion of hydrodynamic effects can be effectively done via transport coefficients~\cite{Supp} in a modified form. For
example, it has been shown recently that complex hydrodynamic effects, such as diffusion in complex liquids,
can be correctly accounted for with wave vector dependent viscosities~\cite{C9SM01119F}.

We expect the same type of superadiabatic forces to be present in other dynamical systems,
including granular systems and inertial Newtonian dynamics such as e.g. molecular liquids.
Within the Navier-Stokes equations one assumes (i) the viscous term to be proportional to
the velocity gradient and (ii) omits the structural terms.
Our results indicate both assumptions might not be acceptable even in very simple flows. 
We anticipate, e.g.\ that the structural forces play a crucial role in most nonequilibrium situations,
including e.g., turbulent flows~\cite{PhysRevX.9.041006}, crystallization \cite{tan2014visualizing,golde2016correlation}, and the
hysteresis of the liquid-solid transition in granular systems~\cite{PhysRevX.9.031027}.

\section*{Acknowledgements}
This work is supported by the German Research Foundation (DFG) via project 436306241. We thank Nikolai Jahreis for useful discussions and
verification of the force splitting in a related system.

\end{document}